\documentclass[]{interact}

\usepackage{epstopdf}% To incorporate .eps illustrations using PDFLaTeX, etc.
\usepackage{subfigure}% Support for small, `sub' figures and tables

\usepackage{natbib}% Citation support using natbib.sty
\bibpunct[, ]{(}{)}{;}{a}{}{,}% Citation support using natbib.sty
\theoremstyle{plain}% Theorem-like structures provided by amsthm.sty

\theoremstyle{definition}

\theoremstyle{remark}

\providecommand{\tightlist}{%
  \setlength{\itemsep}{0pt}\setlength{\parskip}{0pt}}

\usepackage{hyperref}
\usepackage[utf8]{inputenc}
\def\tightlist{}
\usepackage{setspace}
\usepackage{graphicx}
\usepackage{nicematrix}
\NiceMatrixOptions{code-for-first-row = \color{red} ,code-for-last-row = \color{red} ,code-for-first-col = \color{blue} ,code-for-last-col = \color{blue}}

\begin{document}

\articletype{}

\title{New and simplified manual controls for projection and slice
tours, with application to exploring classification boundaries in high
dimensions}

\author{\name{Ursula Laa$^{a}$, Alex Aumann$^{b}$, Dianne
Cook$^{c}$, German Valencia$^{b}$}
\affil{$^{a}$Institute of Statistics, University of Natural Resources
and Life Sciences, Vienna; $^{b}$School of Physics and Astronomy, Monash
University; $^{c}$Department of Econometrics and Business Statistics,
Monash University}
}

\thanks{CONTACT Ursula
Laa. Email: \href{mailto:ursula.laa@boku.ac.at}{\nolinkurl{ursula.laa@boku.ac.at}}, Alex
Aumann. Email: \href{mailto:aaum0002@student.monash.edu}{\nolinkurl{aaum0002@student.monash.edu}}, Dianne
Cook. Email: \href{mailto:dicook@monash.edu}{\nolinkurl{dicook@monash.edu}}, German
Valencia. Email: \href{mailto:german.valencia@monash.edu}{\nolinkurl{german.valencia@monash.edu}}}

\maketitle

\begin{abstract}
This paper describes new user controls for examining high-dimensional
data using low-dimensional linear projections and slices. A user can
interactively change the contribution of a given variable to a
low-dimensional projection, which is useful for exploring the
sensitivity of structure to particular variables. The user can also
interactively shift the center of a slice, for example, to explore how
structure changes in local subspaces. The Mathematica package as well as
example notebooks are provided, which contain functions enabling the
user to experiment with these new manual controls, with one specifically
for exploring regions and boundaries produced by classification models.
The advantage of Mathematica is its linear algebra capabilities, and
interactive cursor location controls. Some limited implementation has
also been made available in the R package \tt{tourr}. 
\end{abstract}

\begin{keywords}
data visualisation; grand tour; statistical computing; statistical
graphics; multivariate data; dynamic graphics
\end{keywords}

\hypertarget{introduction}{%
\section{Introduction}\label{introduction}}

From a statistical perspective 3D is a rare data dimension, so unlike in
most 3D rotation computer graphics applications, the more useful methods
for data analysis need to work for arbitrary dimension. A good approach
is to show projections from an arbitrary dimensional space to create
dynamic data visualizations called \emph{tours}. Tours involve views of
high-dimensional (\(p\)) data with low-dimensional (\(d\)) projections.
In his original paper on the grand tour, \citet{As85} provided several
algorithms for tour paths that could theoretically show the viewer the
data \emph{from all sides}. Prior to Asimov's work, there were numerous
preparatory developments including \citet{tukey}'s PRIM-9. PRIM-9 had
user-controlled rotations on coordinate axes, allowing one to manually
tour through low-dimensional projections. (A video illustrating the
capabilities is available through video library of \citet{ASA22}.)
Steering through all possible projections is impossible, unlike Asimov's
tours which allows one to quickly see many, many different projections.
After Asimov there have been many tour developments, which are
summarized in \citet{lee2021}.

One such direction of work develops the ideas from PRIM-9, to provide
manual control of a tour. \citet{cook_manual_1997} describe controls for
1D (or 2D) projections, respectively in a 2D (or 3D) manipulation space,
allowing the user to select any variable axis, and rotate it into, or
out of, or around the projection through horizontal, vertical, oblique,
radial or angular changes in value. \citet{spyrison_spinifex_2020}
refined this algorithm and implemented it to generate radial tour
animation sequences.

Manual controls are especially useful for assessing sensitivity of
structure to particular elements of the projection. There are many
places where it is useful. In exploratory data analysis, where one sees
clusters in a projection, one may ask whether some variables can be
removed from the projection without affecting the clustering. For
interpreting models, one can reduce or increase a variable's
contribution to examine the variable importance. Having the user
interact with a projection is extremely valuable for understanding
high-dimensional data. However, these algorithms have two problems: (1)
the pre-processing of creating a manipulation space overly complicates
the algorithm, (2) extending to higher dimensional control is difficult.

Another potentially useful manual control, is to allow the user to
choose the position of the center of a slice. The slice tour was
introduced in \citet{slicetour}. It operates by converting the
projection plane into a slice, by removing or de-emphasizing points that
are further than a fixed orthogonal distance from the plane. The
projection plane is usually thought of as passing through the center of
the data. Manual control would allow the user to change the position of
the center point, by shifting it along a coordinate axis, while keeping
the orientation of the projection plane fixed. The purpose would be to
explore how or if the shape of the data, in the space orthogonal to the
projection, changes as one gets away from the center. It would also
allow the user to interactively decide on the thickness of the slice.

This paper explains the new manual controls for projection and slice
tours. The next section describes the new algorithm for manual control,
for both projections and slices. The use of these methods is illustrated
to compare and contrast boundaries constructed by different classifiers.
The software section describes a mathematica package that is used for
the application, and describes the interactive environment that would be
desirable within R as new technology becomes available. The paper is
accompanied by an appendix with more details and adjustments to the
manual controls, and three Mathematica notebooks that can be used to
reproduce the application.

\hypertarget{sec:method}{%
\section{How to construct a manual tour}\label{sec:method}}

A manual tour allows the user to alter the coefficients of one (or more)
variables contributing to a \(d\) dimensional projection. The initial
ingredients are an orthonormal basis (\(A_{p\times d}\)) defining the
projection of the data, and a variable id (\(m \in \{1, ..., p\}\))
specifying which coefficient will be changed. A method to update the
values of the component (\(m^{th}\) row of \(A_{p\times d}\)) of the
controlled variable \(V_m\) is then needed.

\hypertarget{existing-methods}{%
\subsection{Existing methods}\label{existing-methods}}

The methods for updating component values in \citet{cook_manual_1997}
(and utilized in \citet{spyrison_spinifex_2020}) are prescribed
primarily for a 2D projection, to take advantage of (then) newly
developed 3D trackball controls made available for computer gaming. The
first step was to construct a 3D manipulation space from a 2D
projection. In this space, the coefficient of the controlled variable
ranges between -1 and 1. Movements of a cursor are recorded and
converted into changes in the values of \(V_m\) thus changing the
displayed 2D projection. The algorithm also provided constraints to
horizontal, vertical, radial or angular motions only. The construction
of the manipulation space overly complicates the manual controls,
especially when considering possible techniques that will apply to
arbitrary \(d\).

\hypertarget{a-new-simpler-and-broadly-applicable-approach}{%
\subsection{A new simpler and broadly applicable
approach}\label{a-new-simpler-and-broadly-applicable-approach}}

The new approach emerged from experiments on the tour using the linear
algebra capabilities, and relatively new interactive graphics interface,
available in Mathematica \citep{Mathematica}. The components
corresponding to \(V_m\) are directly controlled by cursor movement,
which updates row \(m\) of \(A\). The updated matrix is then
orthonormalised.

\hypertarget{algorithm}{%
\subsubsection{Algorithm}\label{algorithm}}

\begin{enumerate}
\def\labelenumi{\arabic{enumi}.}
\tightlist
\item
  Provide \(A\), and \(m\). (Note that \(m\) could also be automatically
  chosen as the component that is closest to the cursor position.)
\item
  Change values in row \(m\), for example, if \(d=2\) gives \[
  A^* = [ \boldsymbol{a}^*_1~\boldsymbol{a}^*_2 ] = \left[ \begin{array}{cc} a_{11} & a_{12}\\
                             \vdots & \vdots \\
                             a^*_{m1} & a^*_{m2}\\
                             \vdots & \vdots \\
                             a_{p1} & a_{p2} 
       \end{array}\right].
  \] \noindent A large change in these values would correspond to making
  a large jump from the current projection. Small changes would
  correspond to tracking a cursor, making small jumps from the current
  projection.
\item
  Orthonormalise \(A^*\), using Gram-Schmidt.

  \begin{enumerate}
  \def\labelenumii{\roman{enumii}.}
  \tightlist
  \item
    Normalise \(\boldsymbol{a}^*_1\) and \(\boldsymbol{a}^*_2\).
  \item
    \(\boldsymbol{a}^*_2 = \boldsymbol{a}^*_2 - (\boldsymbol{a}^*_1\cdot\boldsymbol{a}^*_2)\boldsymbol{a}^*_1\).
  \end{enumerate}
\end{enumerate}

\begin{figure}
\includegraphics[width=1\linewidth]{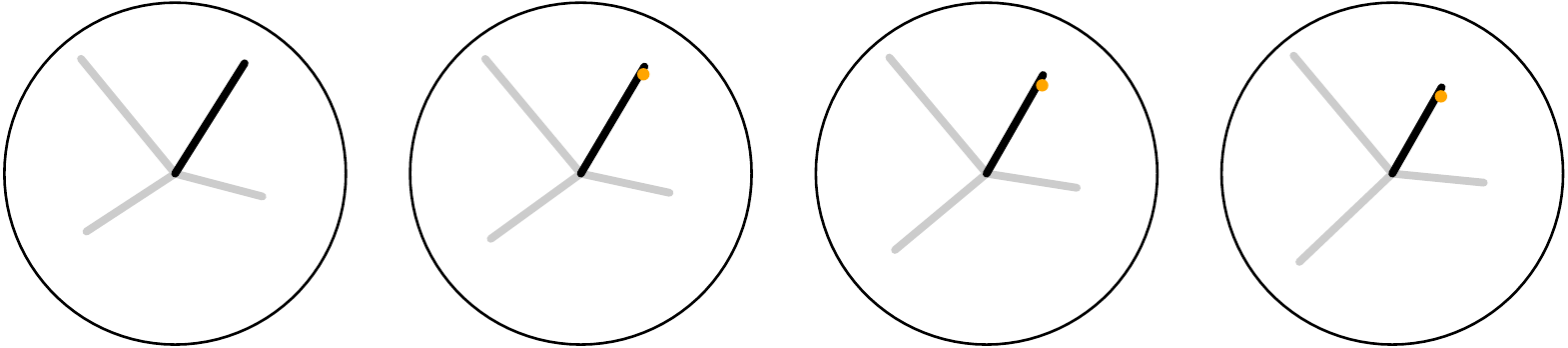} \caption{Sequence of projections where contribution of one variable is controlled (black) is changed using unconstrained orthonormalisation. The dot (orange) indicates the chosen values for the controlled variable. It can be seen that the actual axis does not precisely match the chosen position, but it is close.}\label{fig:manualsequence}
\end{figure}

This algorithm will produce the changes to a projection as illustrated
in Figure \ref{fig:manualsequence}. The controlled variable, \(V_m\),
corresponds to the black line, and sequential changes to row \(m\) of
\(A\) can be seen to roughly follow a specified position (orange dot).
Changes in the other components happen as a result of the
orthonormalisation, but are uncontrolled.

\hypertarget{refinements-to-enforce-exact-position}{%
\subsection{Refinements to enforce exact
position}\label{refinements-to-enforce-exact-position}}

The problem with the new simple method is that it is not faithful to the
precise values for \(V_m\) because the orthonormalisation will change
them. Even though these changes are for the most part imperceptible, one
may wish to avoid them and there are numerous ways that this can be
enforced, a few are detailed in the Appendix. These primarily differ in
how the remaining variables are adjusted during orthonormalisation.

\hypertarget{manual-control-for-slices}{%
\subsection{Manual control for slices}\label{manual-control-for-slices}}

To better explore the space we combine the manual controls for the
projection with manual controls for slicing. A slice is a section of the
data that is defined by a projection, a center point that is anchoring
it in the high-dimensional space and the slice thickness \(h\)
\citep{slicetour}. A data point is inside the slice if its orthogonal
distance from the projection plane (passing through the center point) is
below the thickness \(h\). This orthogonal distance is computed in terms
of the component that is normal on the projection plane. For
\(\mathbf{x}_i\) a \(p\) dimensional data point and \(\mathbf{c}\) the
center point (in the same \(p\) dimensional space) we compute the
orthogonal distance as \begin{equation}
v_i^2 = ||\mathbf{x}_i' - \mathbf{c}'||^2 = \mathbf{x}_i'^2 + \mathbf{c}'^2 - 2 \mathbf{x}_i'\cdot\mathbf{c}' ,
\label{eq:slice}
\end{equation} with
\(\mathbf{c}' = \mathbf{c} - (\mathbf{c}\cdot \mathbf{a}_1) \mathbf{a}_1 - (\mathbf{c}\cdot \mathbf{a}_2 )\mathbf{a}_2\),
\(\mathbf{x}_i' = \mathbf{x}_i - (\mathbf{x}_i\cdot \mathbf{a}_1) \mathbf{a}_1 - (\mathbf{x}_i\cdot \mathbf{a}_2) \mathbf{a}_2\)
and \(\mathbf{a}_k, k=1,2 (=d)\) denoting the columns of the projection
matrix, \(\mathbf{A}=(\mathbf{a}_1, \mathbf{a}_2)\).

\hypertarget{shifting-the-center}{%
\subsubsection{Shifting the center}\label{shifting-the-center}}

A natural starting point is to place \(\mathbf{c}\) in the center of the
data distribution, but shifting it away from the mean can provide
additional insights. In the case of a single orthogonal direction on the
projection plane we can pick a sequence of center points \(\mathbf{c}\)
in steps along that direction to move the slice and fully cover the data
space. This no longer works in higher-dimensional spaces, and we can
think of picking one direction and shifting the slice along the
component orthogonal to the projection plane.

\hypertarget{changing-the-thickness}{%
\subsubsection{Changing the thickness}\label{changing-the-thickness}}

In addition it is also useful to interactively change the slice
thickness \(h\) (also called the slice radius), in particular to find
the preferred value for exploring the input data. For guidance the
estimates of the number of points inside the slice as a function of the
original sample size \(N\) and the number of dimensions \(p\) from
\citet{sectionpursuit} can be used: in case of a uniform distribution
inside a sphere of radius \(R\) a slice with thickness \(h\) will
contain \(N_S\) points, with \begin{equation}
N_S(h, p, R, N) = \frac{N}{2} \left(\frac{h}{R}\right)^{p-2} \left(p - (p-2)\left(\frac{h}{R}\right)^{2}\right).
\label{eq:count}
\end{equation}

\hypertarget{sec:implementation}{%
\section{Software}\label{sec:implementation}}

The implementation of the manual tour as suggested here requires the
visualization of the current projection in terms of an axis display (see
Figure \ref{fig:manualsequence} for an example). This display should be
able to track the mouse position and adjust the projection based on the
user selection. The user can select one axis (corresponding to one
variable) by clicking on it, and then adjust the position of that axis
by dragging it during the click event. In practice the closest axis will
be selected, and the dragging results in tracking of the mouse position
which will update the projection in small steps, such that no further
interpolation is required.

The interactions in the axis display need to be mapped back onto the
current projection matrix, which will then be orthonormalized before
feeding back into the axis display. A second, linked display shows the
projected data in sync with the updates from the axis display.

In addition, we also want to be able to look at slices of the data and
select slicing parameters interactively. Here we have implemented a
switch to change between projection and slice view, a slider that
adjusts the slice thickness \(h\) (dubbed \texttt{height} in the
\texttt{Mathematica} package) and a numeric input to specify the center
point \(c\) explicitly. The display will update based on these inputs,
in particular the view will jump to a new slice when \(c\) is changed
(though an interpolation might also be useful).

One might implement such an interactive interface via R Shiny, but in
particular the tracking of the mouse position in small increments might
pose a challenge. Here we found the dynamic graphics interface from
Mathematica to be especially useful. In the following we first describe
the relevant Mathematica functionalities and how they were used in the
implementation, followed by a brief sketch of how the new approaches are
implemented in the \texttt{tourr} package. There is hope that better
interactivity will be available for R soon, which might allow for
implementation of the techniques developed in Mathematica.

\hypertarget{mathematica-package}{%
\subsection{Mathematica package}\label{mathematica-package}}

Mathematica provides much utility and versatility, such as inbuilt data
visualization, data manipulation and analysis, dynamic functionality,
and symbolic and numeric computation. Most of this inbuilt functionality
is user-friendly and described in the Mathematica documentation;
typically, numerous examples are provided within the documentation, and
some possible issues are outlined there.

Importantly for our work, it is relatively simple to create dynamic
objects via the inbuilt commands \texttt{Manipulate} or
\texttt{DynamicModule}. Control objects, such as sliders, locator panes,
and input fields, can then be used on dynamic variables, and when there
is a change in the dynamic variable the dynamic objects which contain
that variable will be updated. This is the essential ingredient for this
implementation of the manual tour.

The most relevant inbuilt Mathematica function for our purpose is the
\texttt{LocatorPane}. This creates a region on the screen where the
position of the mouse is captured and then converted to input that
updates the graphics functions, enabling the manual navigation of the
tour.

The examples presented in the application section below (and attached in
notebook format in the supplementary material) all use our primary new
function, \texttt{SliceDynamic}. This function typically accepts grouped
data in the form of a matrix where the second last column details the
name of the group and the last column details the group index (which can
be one if there is only one group). After specifying the initial slice
thickness and the slice range, the user is presented with an interactive
display in which the control objects appear on the left and the slice
visualization appears on the right. The user can change the orientation
of the slice via the locator pane, which changes the projection matrix;
a slider controls the slice thickness; and there is an input field,
which changes the slice center. The user can also change the appearance
of the plot by zooming into the center or changing the point size with
the sliders provided. It is worth noting that this zoom works best with
data that has been centered and scaled. The projected data can be
displayed to contrast it with the slice by ticking a box. The explicit
projection matrix can also be displayed via another checkbox. This is
especially useful when we wish to import a projection that was
identified as interesting in the manual exploration into a later study,
as shown in the applications below.

Details about the implementation and usage instructions are given in the
Appendix.

\hypertarget{extensions-to-the-r-package-tourr}{%
\subsection{\texorpdfstring{Extensions to the R package
\texttt{tourr}}{Extensions to the R package tourr}}\label{extensions-to-the-r-package-tourr}}

The R package tourr \citep{tourr} provides numerous types of tours. An
additional function, \texttt{radial\_tour()} has been constructed that
will generate a sequence of projections that will decrease the
coordinates for \(V_m\) to \(\boldsymbol{0}\) and back to the original
values. It is applicable for any projection dimension, \(d\). Changing
slice center manually can be accomplished with the new function,
\texttt{manual\_slice()}. This changes the value of the center point
(\(\mathbf{c}\)) of the slice, along a selected variable axis. It slices
from the center of the data out to one side, back to the center, out to
the opposite side, and then back to the center. The appendix contains a
discussion of a possible visual guide for the slice center location.

\hypertarget{sec:examples}{%
\section{Application}\label{sec:examples}}

\begin{figure}

{\centering \includegraphics[width=0.6\linewidth]{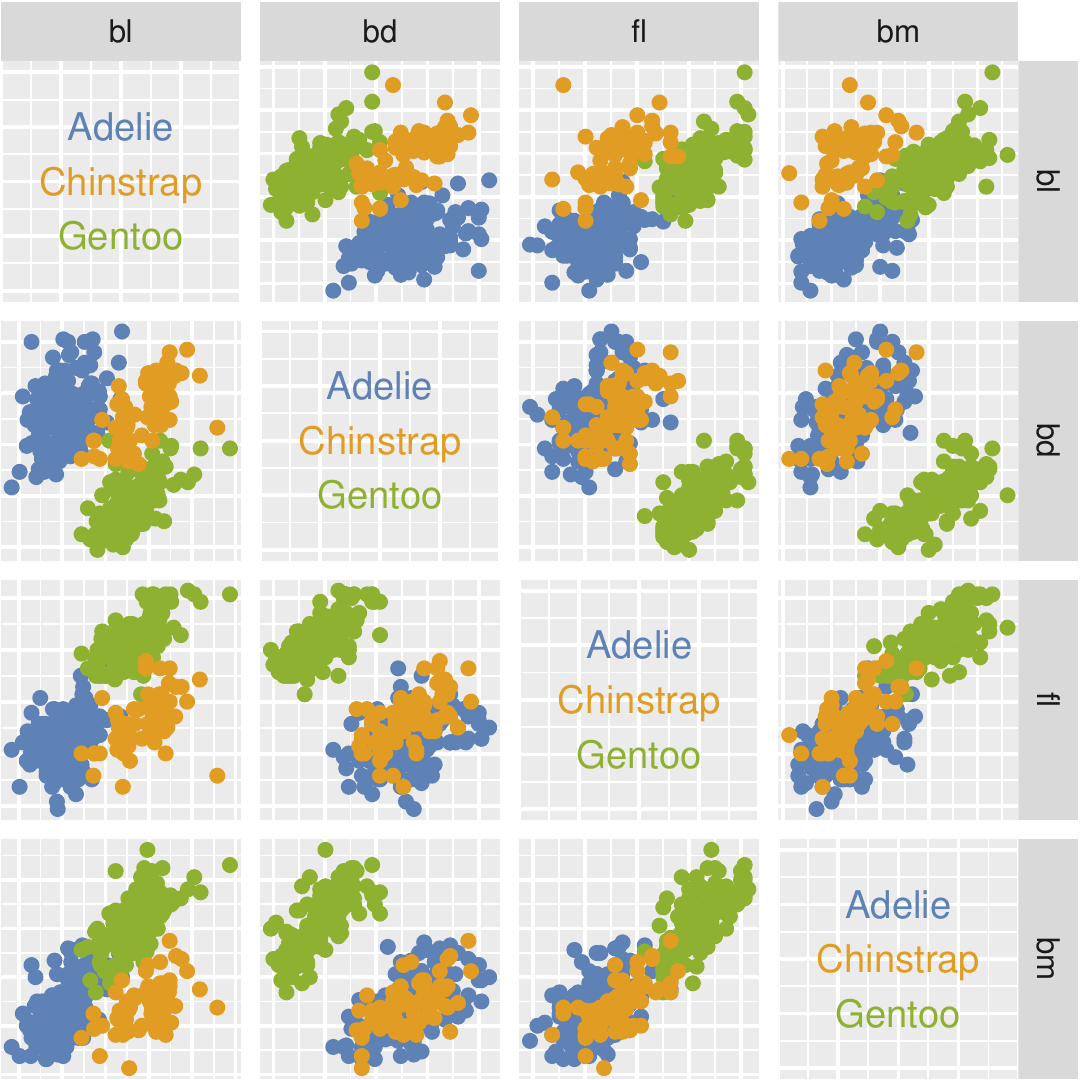} 

}

\caption{Scatterplot matrix of the (standardised) penguins data. The three species are reasonably different in size, with Gentoo distinguised from the other two on body depth relative to flipper length and body mass.}\label{fig:penguins-scatmat}
\end{figure}

To illustrate the usefulness of the manual controls we use the 4D
penguins data \citep{penguins} and look at classification models
following \citet{sam11271}. We will show how classification boundaries
can be explored and better understood on projections and slices through
4D space. Figure \ref{fig:penguins-scatmat} shows a scatterplot matrix
of this data. There are four variables
(\texttt{bl\ =\ bill\_length\_mm,\ bd\ =\ bill\_depth\_mm,\ fl\ =\ flipper\_length\_mm,\ bm\ =\ body\_mass\_g})
measuring the size of the penguins from three species (Adelie, Chinstrap
and Gentoo). The scatterplot matrix shows that the three species appear
to be likely separable, and that at least the Gentoo can be
distinguished from the other two species when \texttt{bd} is paired with
\texttt{fl} or \texttt{bm}. The steps for exploring boundaries in this
example are as follows:

\begin{enumerate}
\def\labelenumi{\arabic{enumi}.}
\tightlist
\item
  Build your classification model.
\item
  Predict the class for a dense grid of values covering the data space.
\item
  Examine projections, using a manual tour so that the contribution of
  any variable is controlled.
\item
  Slice through the center, to explore where the boundaries will likely
  meet.
\item
  Move the slice by changing the center in the direction of a single
  variable to explore the extent of a boundary for a single group
  relative to a variable.
\end{enumerate}

\hypertarget{constructing-the-4d-prediction-regions}{%
\subsection{Constructing the 4D prediction
regions}\label{constructing-the-4d-prediction-regions}}

We use the \texttt{classifly} package \citep{classifly} to generate
predictions across the 4D cube spanned by the data, with two
classification models: linear discriminant analysis (LDA) and random
forest (RF). Both the data and the model points in the grid are centered
and scaled (standard deviation \(= 1\)).

\hypertarget{exploring-projections-manually}{%
\subsection{Exploring projections
manually}\label{exploring-projections-manually}}

\begin{figure*}[ht]
\centerline{\includegraphics[width=0.32\textwidth]{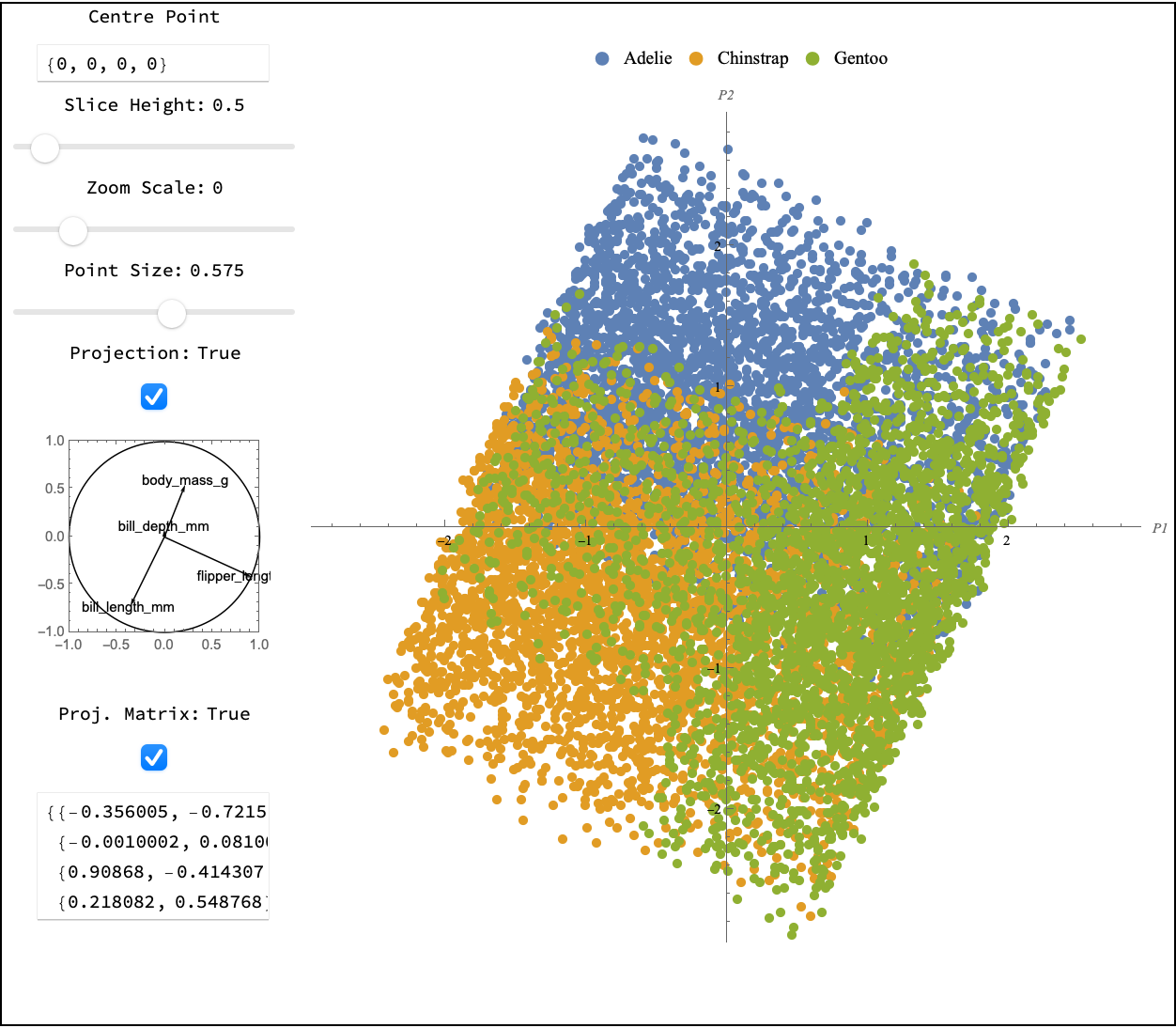}
\includegraphics[width=0.32\textwidth]{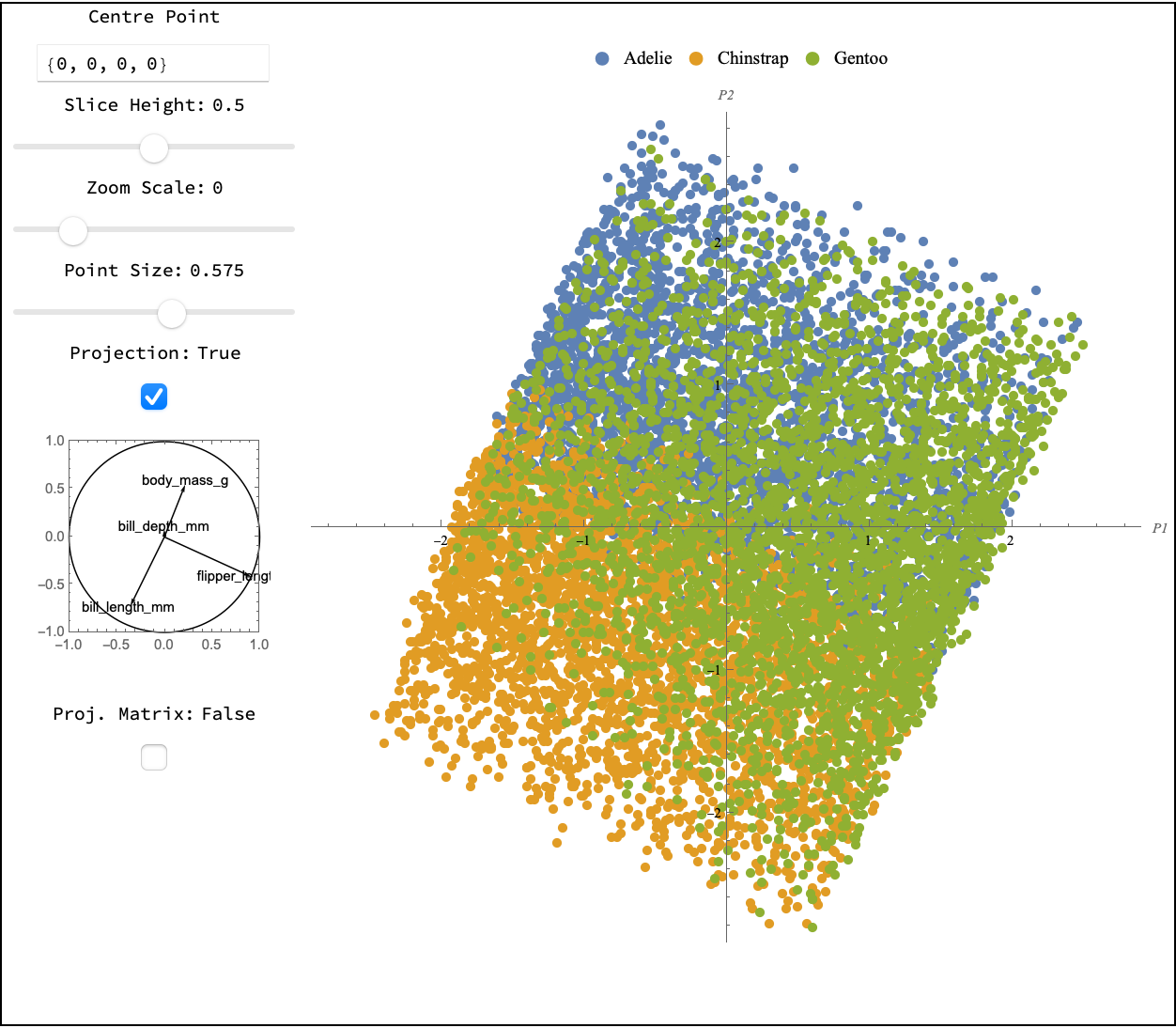}
\includegraphics[width=0.32\textwidth]{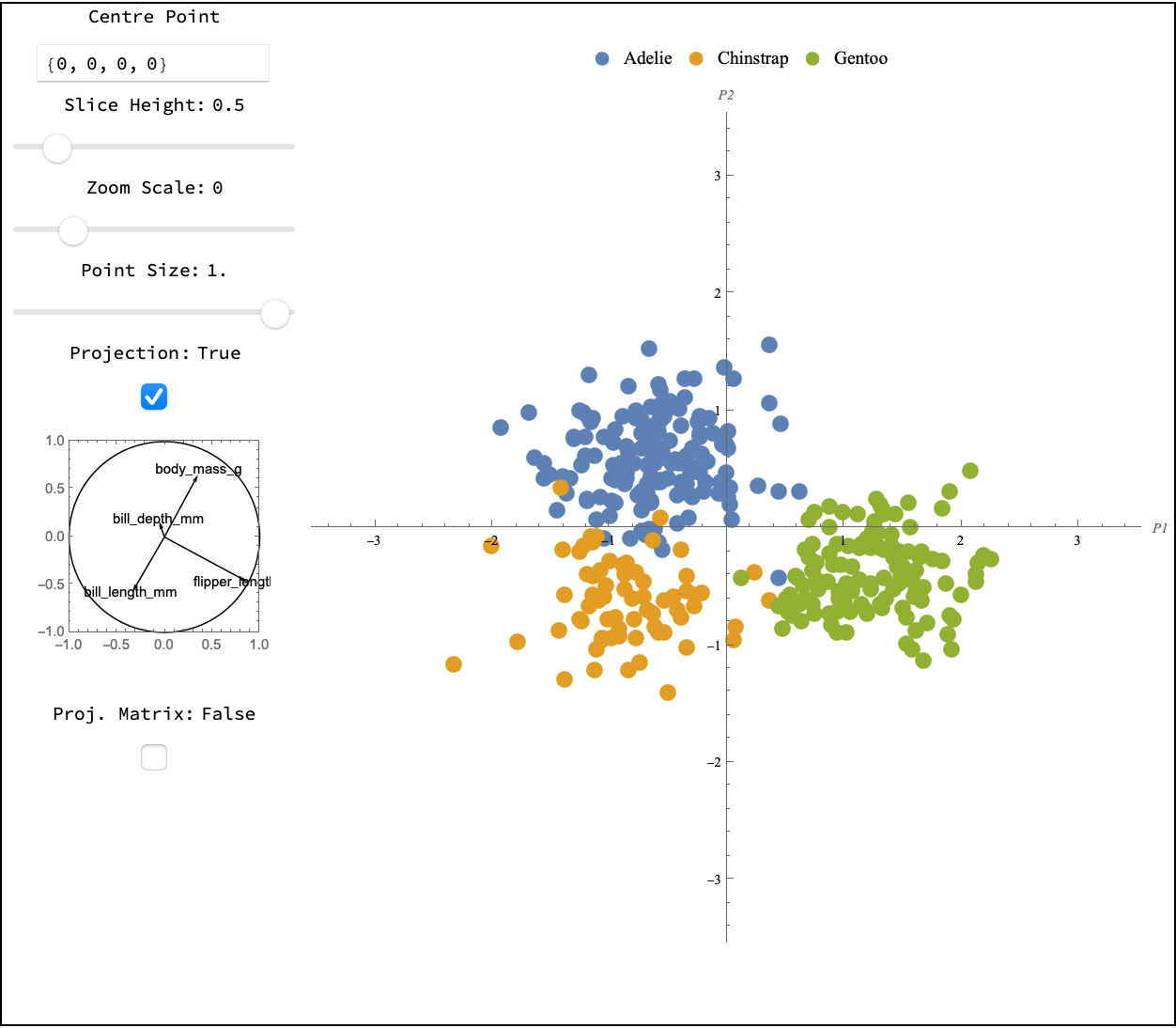}}
\caption{Projection identified using the manual tour, because it reveals an interesting structure in the predictions from the RF model (left). We can clearly see a block structure, while the LDA model (middle) produces linear boundaries. The three groups are nicely separated in this projection of the data (right).}
\label{proj1}
\end{figure*}

We start by exploring the projections of the model prediction. Figure
\ref{proj1} summarizes the process. Rotating manually the view, we can
visualize the location of predictions for each of the three species, and
also get a sense of the difference between the two models. To illustrate
this difference, we have manually rotated the projection for the RF
model (left plot) to identify a view that shows the non-linear but
block-type structure that is typical of this type of model. This
particular projection (\(A_1\)) is exported so that it can later be used
to show the LDA model (middle plot) and the actual data (right plot).
What can be seen is the linearity of the LDA model, where the boundaries
are linear and oblique to the variable axes. And, interestingly, this
particular projection of the original data shows very distinct clusters
of the three species. That means, the obscuring of the boundaries
between groups for both of the models is driven by what is happening in
the orthogonal space to the plane of the selected projection.

This example can be reproduced with the code in
penguins\_exploring\_manually.nb and a run through is shown in the video
at \url{https://vimeo.com/747585410}.

\hypertarget{slicing-through-the-center}{%
\subsection{Slicing through the
center}\label{slicing-through-the-center}}

We now continue the investigation by slicing orthogonally to the
projection \(A_1\). For both models we look at a thin (\(h=0.5\)) slice
through the center, \(\mathbf{c}^0 = (0,0,0,0)\). At first we explore
how changing the projection away from \(A_1\) can help with
understanding the boundary better. For our example, notice that \(A_1\)
does not contain any contribution from the second variable
(\texttt{bd}), so we will first rotate this variable into the view.

\begin{figure*}[ht]
\centerline{\includegraphics[width=0.45\textwidth]{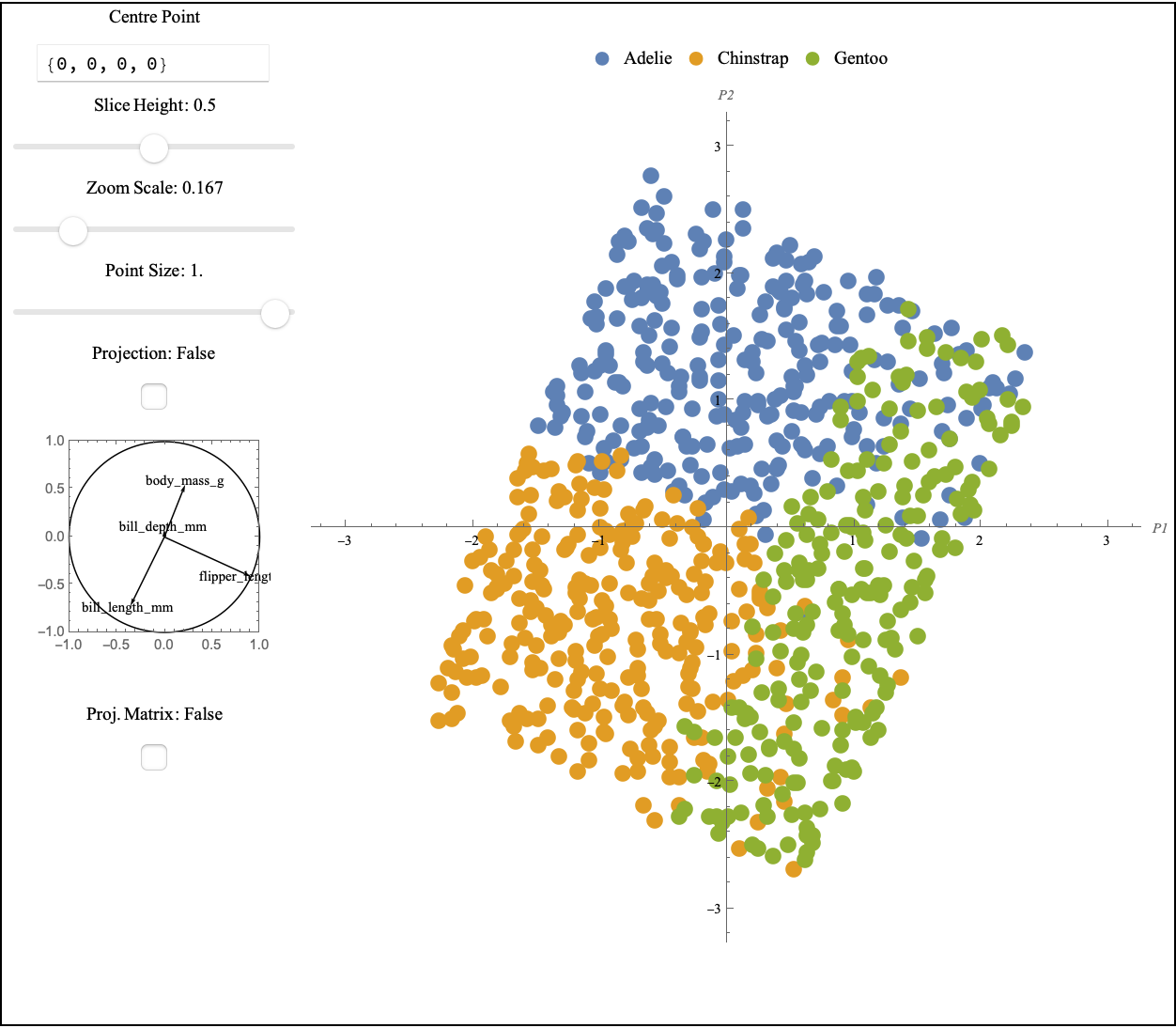}
\includegraphics[width=0.45\textwidth]{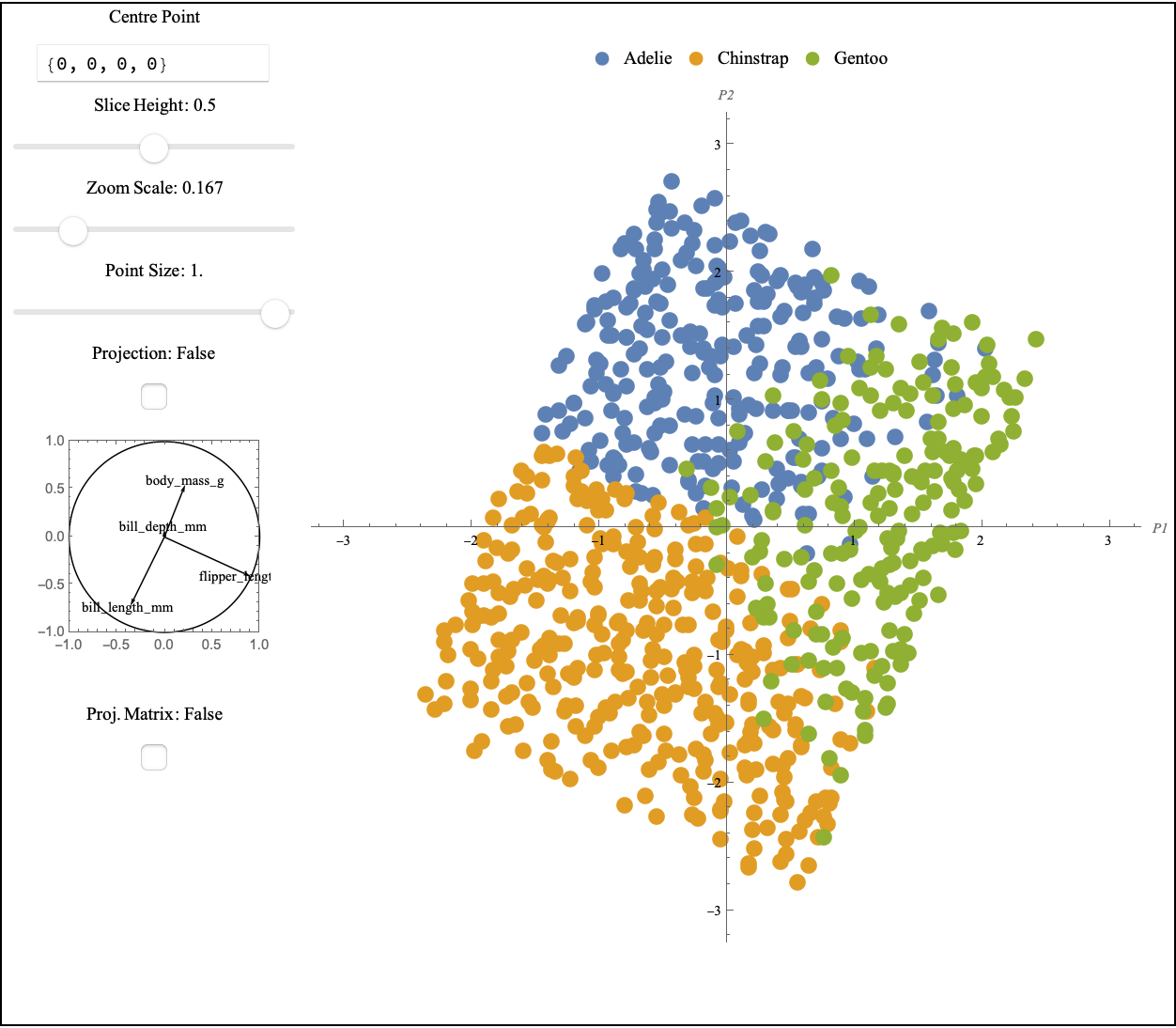}}
\centerline{\includegraphics[width=0.45\textwidth]{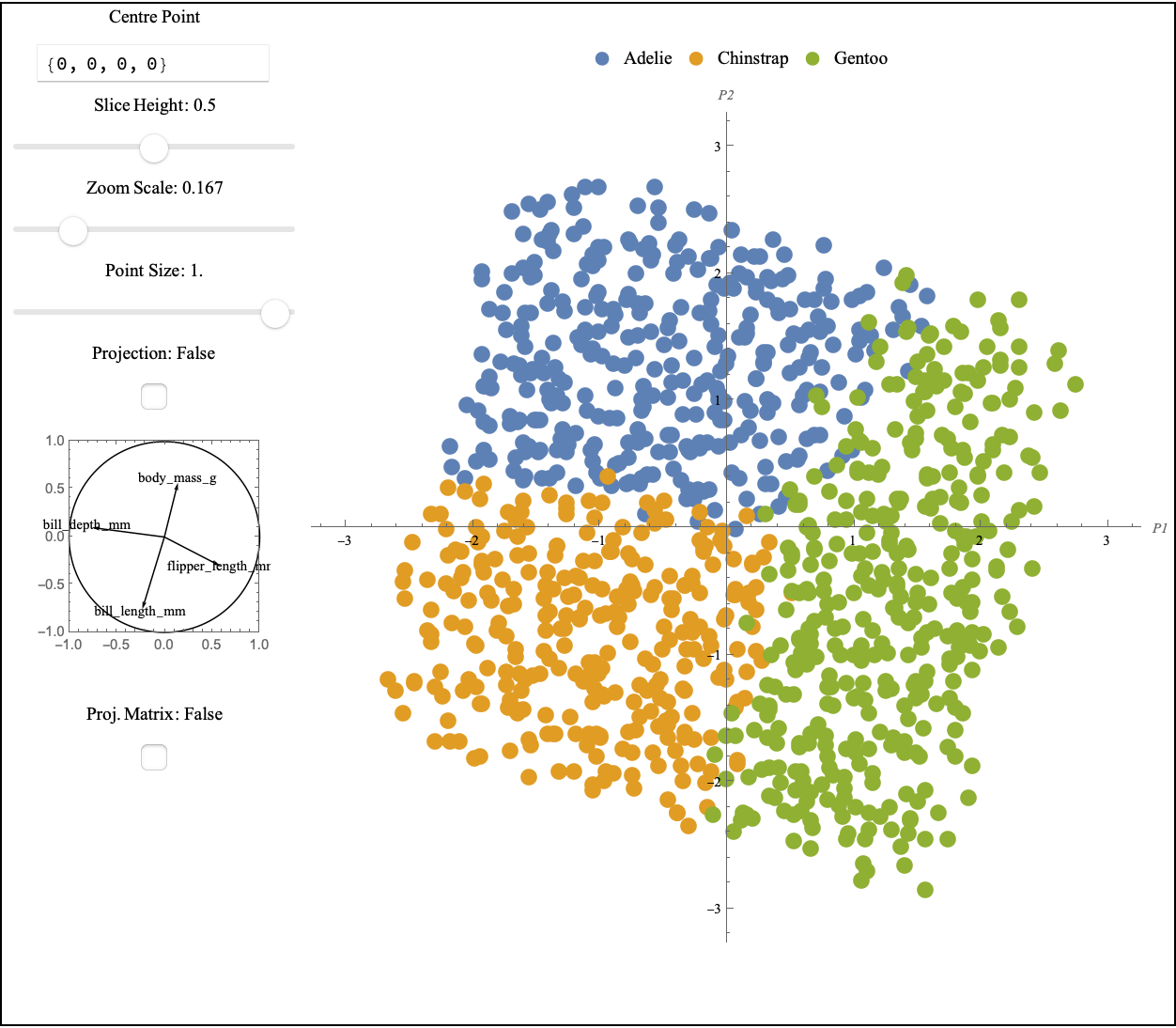}
\includegraphics[width=0.45\textwidth]{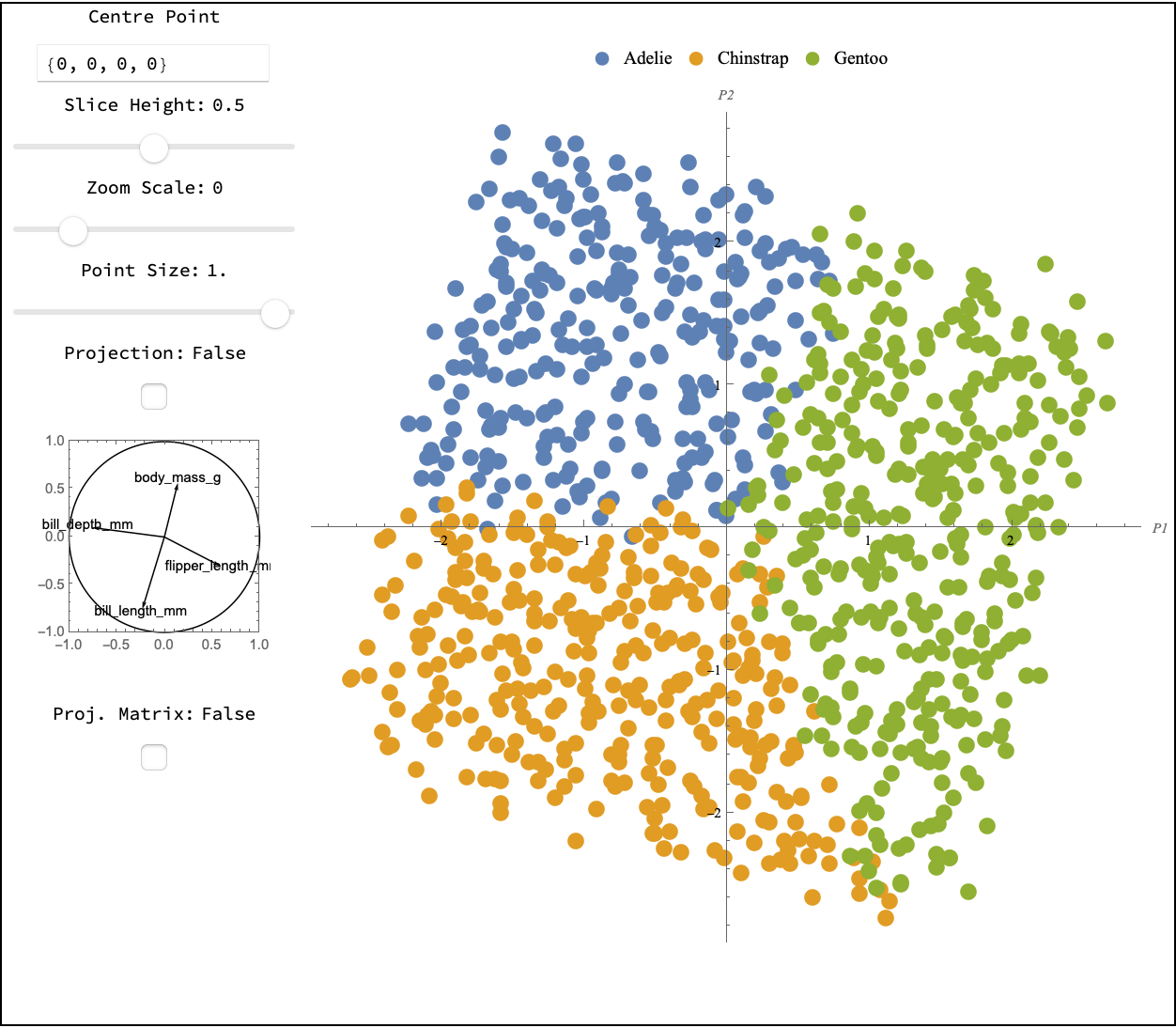}}
\caption{Comparing slices based on two projections $A_1$ (top row) and $A_2$ (bottom row), for the two models RF (left) and LDA (right). With $A_1$ we see two groups overlap (green - Gentoo with yellow - Chinstrap), while the rotation to $A_2$ results in clear boundaries inside the slice. The boundary between Adelie (blue) and Chinstrap (yellow) is similar for both models but very different between Chinstrap and Gentoo (green).}
\label{slice1}
\end{figure*}

Figure \ref{slice1} shows snapshots of the exploration. The top row is
the initial projection (\(A_1\) and \(\mathbf{c}^0\)), and the bottom
row is a later projection containing more of \texttt{bd} (new projection
\(A_2\) while keeping the center fixed at \(\mathbf{c}^0\)). The columns
correspond to the two models, with RF on the left and LDA on the right.

Both models have some overlap at the center for the thin slice based on
\(A_1\) (despite the small value for \(h\)). The second slice (based on
\(A_2\)) mostly resolves this overlap and reveals the primary
differences between the models. The boundary between Adelie (blue) and
Chinstrap (yellow) is very similar for both models. The boundary between
Chinstrap and Gentoo (green) is where they differ.

The RF is almost straight in this view, as something we might expect
from a tree model where splitting occurs on single variables only.
However, it is straight in a combination of the second and third
variable (\texttt{bd} and \texttt{fl}). By examining the scatterplot
matrix, we can understand how this boundary was built: on each of
\texttt{bd} and \texttt{fl} it is possible to cut on a value where most
of the Gentoo penguins are different from the other two species for any
tree, and likely only one is needed. Thus the forest construction is
providing a sample of trees that use one variable or the other variable
to split, producing the blocky boundary on the combination of the two.

In contrast, the LDA boundary has an oblique split between the two, and
roughly divides the space into three similarly sized areas. It is almost
like a textbook illustration of how LDA works for two variables (2D)
where by assuming the clusters have equal variance-covariance, it places
a boundary half-way between the group means.

\begin{figure*}[ht]
\centerline{\includegraphics[width=0.45\textwidth]{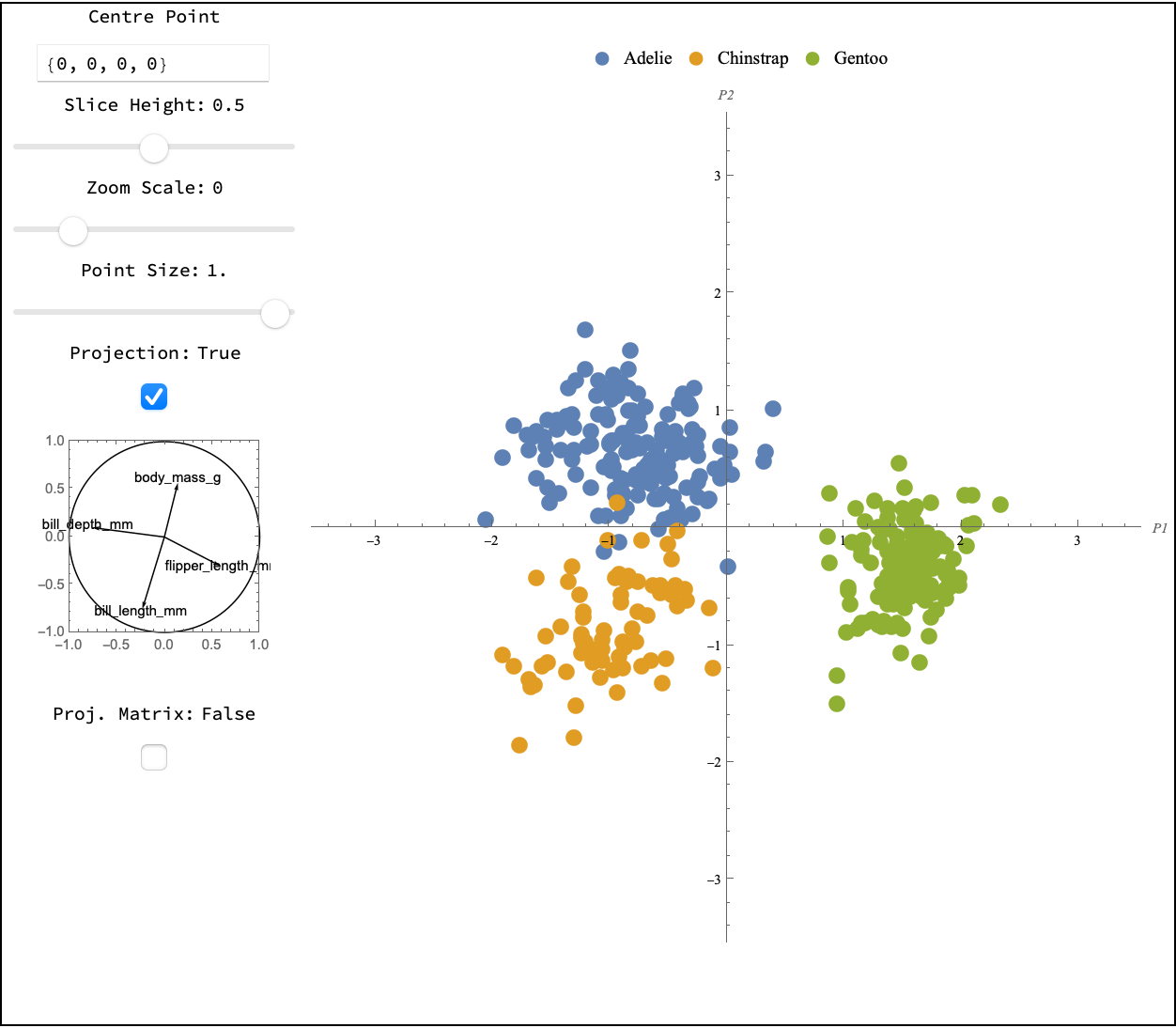}}
\caption{Projection of the data based on $A_2$. Compared to projecting onto $A_1$ we see that the green observations (Gentoo) are more separated from the other two species.}
\label{proj2}
\end{figure*}

Finally, to determine which model describes the boundaries better,
compare them with the \(A_2\) projection of the data (Figure
\ref{proj2}). Gentoo is more separated from the other two in this
projection, and one can imagine that the trees in the forest has
greedily grasped any one of many places to make a split to separate the
group. It might be argued though the that RF boundary is cut too close
to the Chinstrap species, and might lead to some unnecessary
misclassification with new data. The LDA boundary is better placed for
all species.

This example can be reproduced with the code in
penguins\_slicing\_through\_center.nb and a run through is shown in the
first part of the video at \url{https://vimeo.com/747590472}.

\hypertarget{shifting-the-slice-center}{%
\subsection{Shifting the slice center}\label{shifting-the-slice-center}}

We have seen that starting from \(A_1\) using the manual controls to
change the contribution of the second variable we could find a clear
separation boundary indicating the relation between this variable (bill
depth) and the Gentoo penguin species. Instead of rotating to a
different projection, we might also change the view by moving the slice
along one axis in the 4D space. Here we will continue our exploration of
the dependence on \texttt{bd} and move the slice defined by \(A_1\) to
either large positive or negative values (\(c^{\pm}_{bd} = \pm 1.5\)
after centering and scaling). Thus we define
\(\mathbf{c}^{\pm} = (0, c^{\pm}_{bd}, 0, 0)\). Here we will also look
at slices of the observed data points, using a thicker slice (\(h=1.5\))
to capture enough points in a given view.

\begin{figure*}[ht]
\centerline{\includegraphics[width=0.32\textwidth]{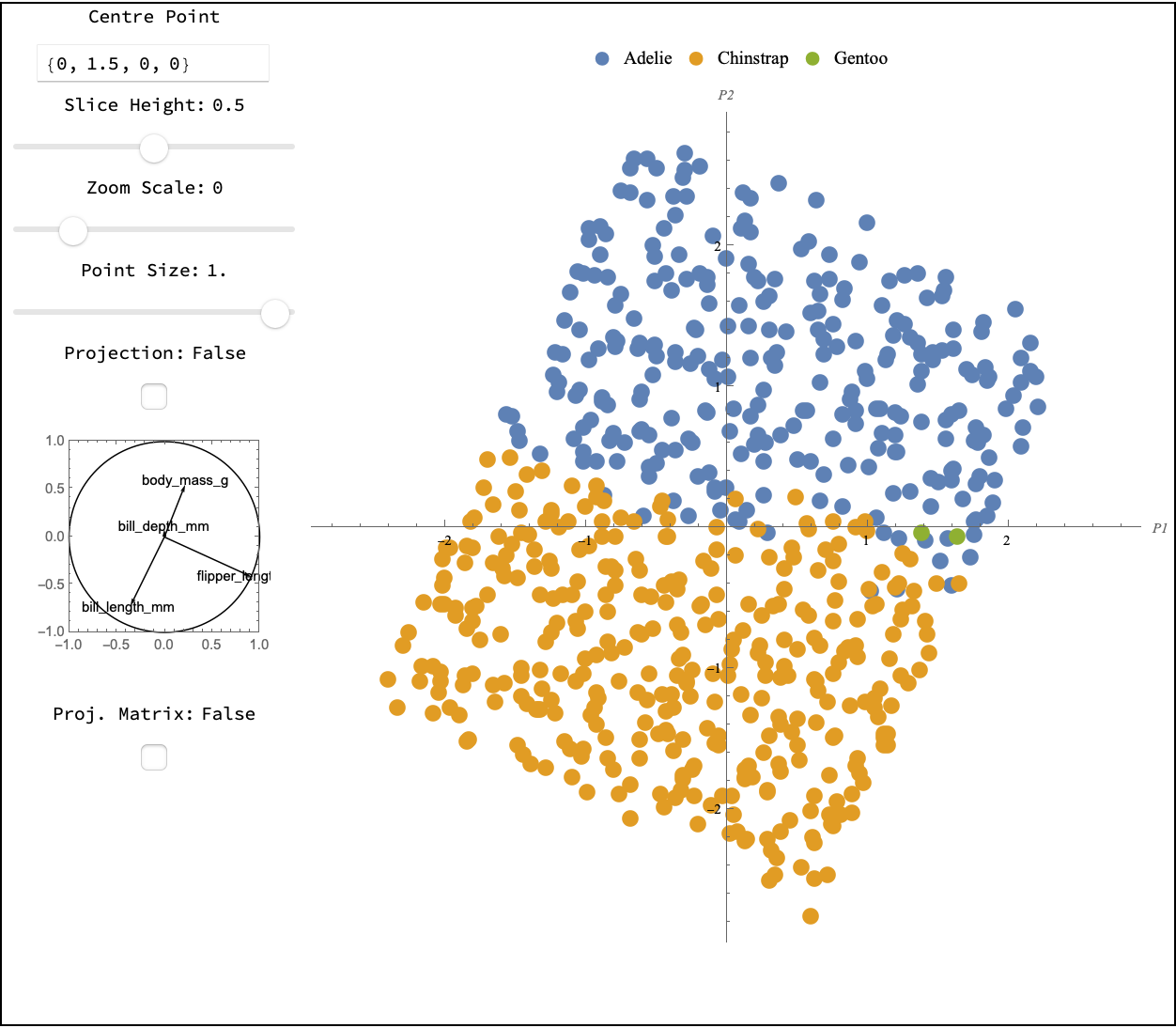}
\includegraphics[width=0.32\textwidth]{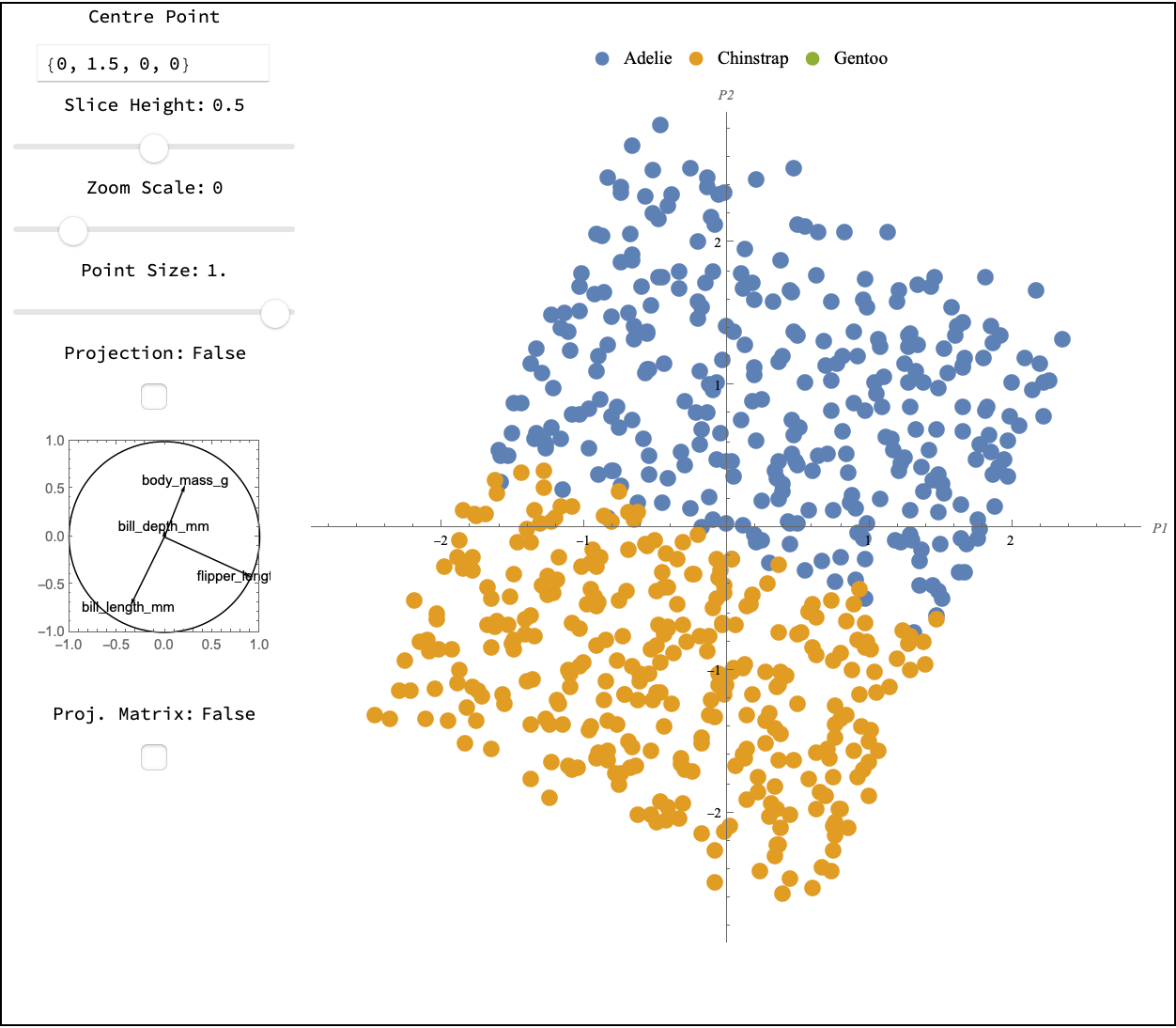}
\includegraphics[width=0.32\textwidth]{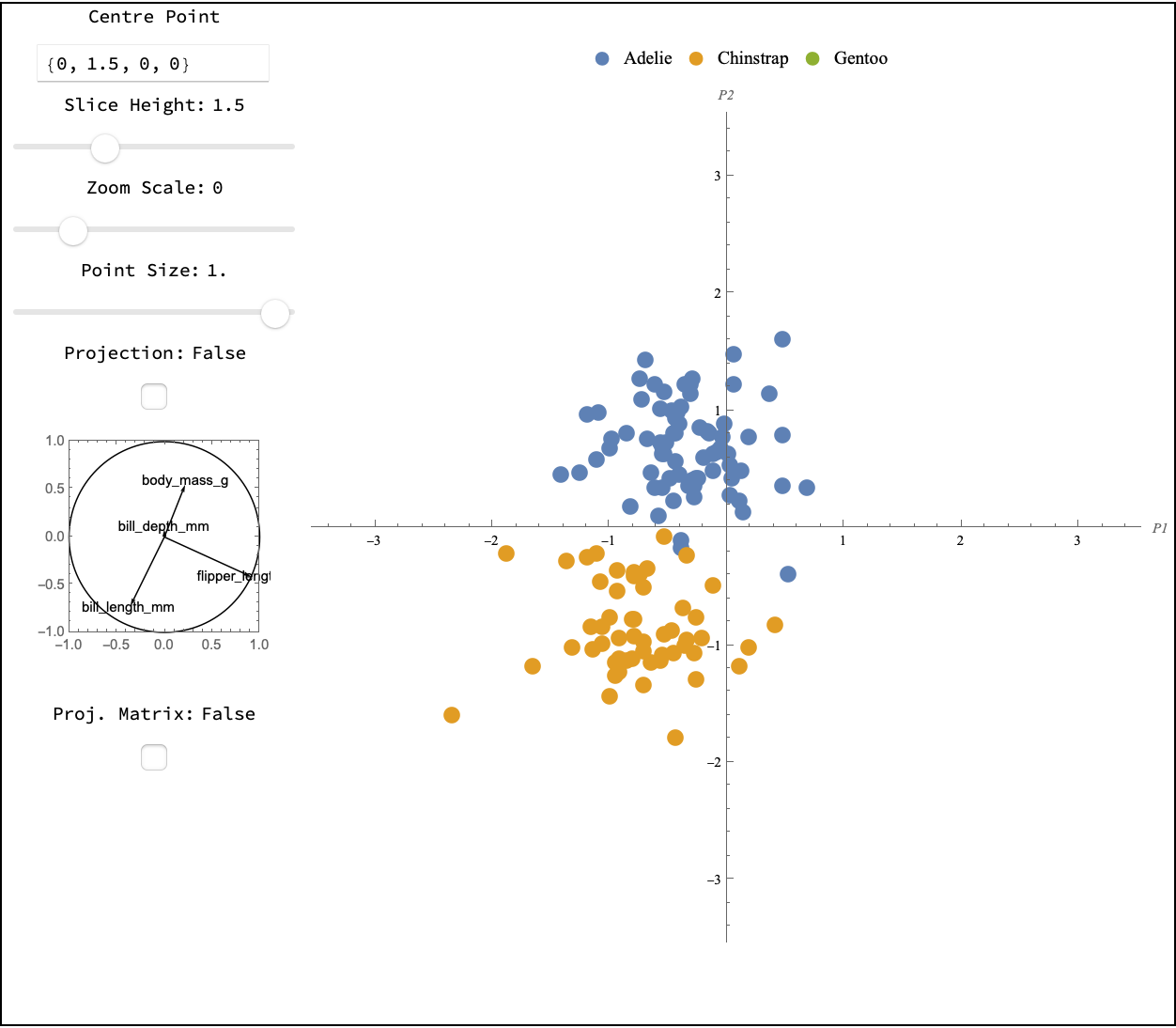}}
\caption{Shifting the slice center in the positive direction of bill depth produces regions that have no Gentoo (green). The two models have similar boundaries except that the RF is more of a step function.}
\label{slice1p}
\end{figure*}

We start by a comparison of the two models and the data distribution
shifting to \(\mathbf{c}^{+}\), thus the slice is localized towards high
values of bill depth in Figure \ref{slice1p}. We can see that all three
slices (the two models and the data) contain almost no points from the
third class (green, Gentoo), and that the decision boundary between the
two models is very similar.

\begin{figure*}[ht]
\centerline{\includegraphics[width=0.32\textwidth]{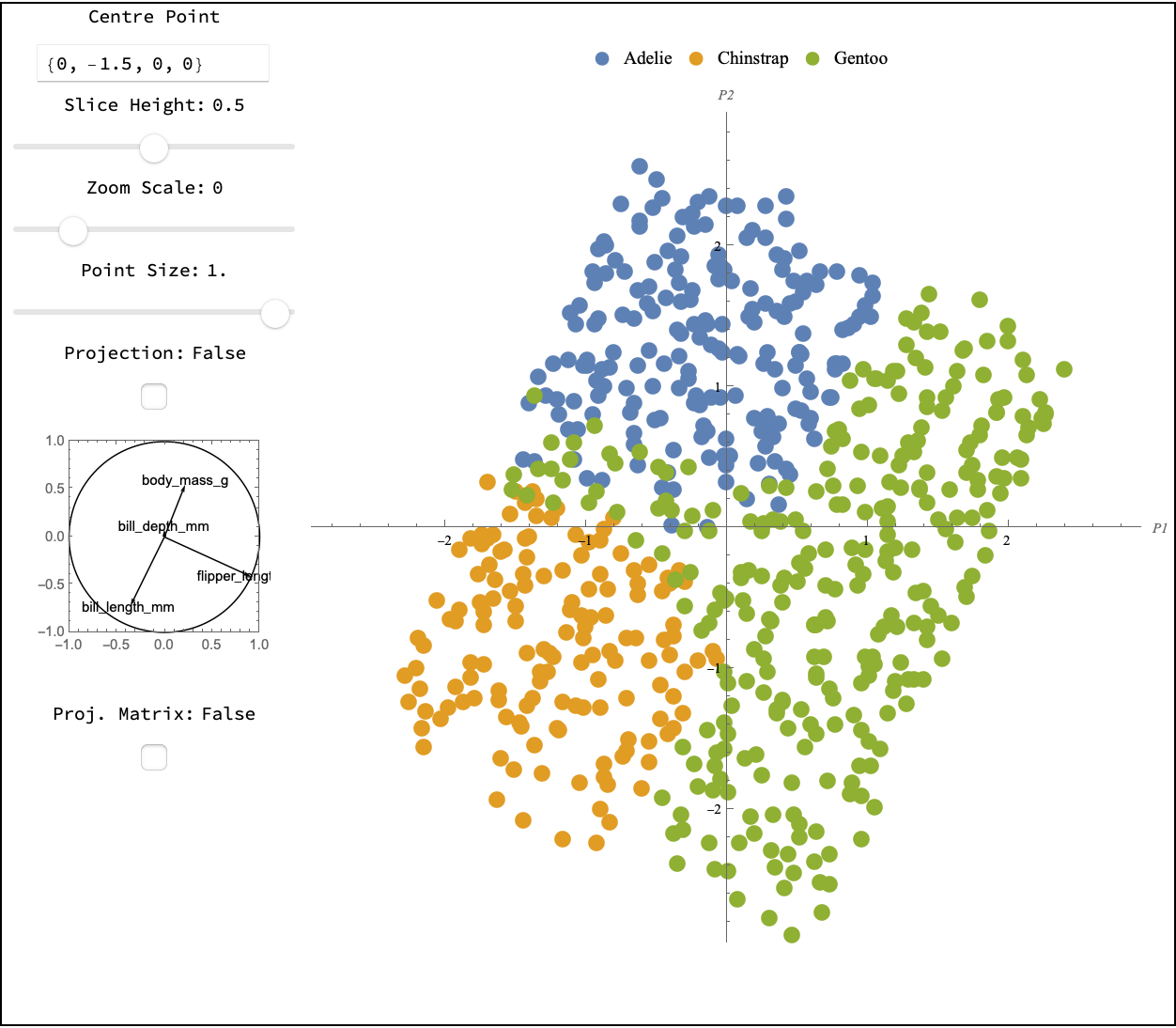}
\includegraphics[width=0.32\textwidth]{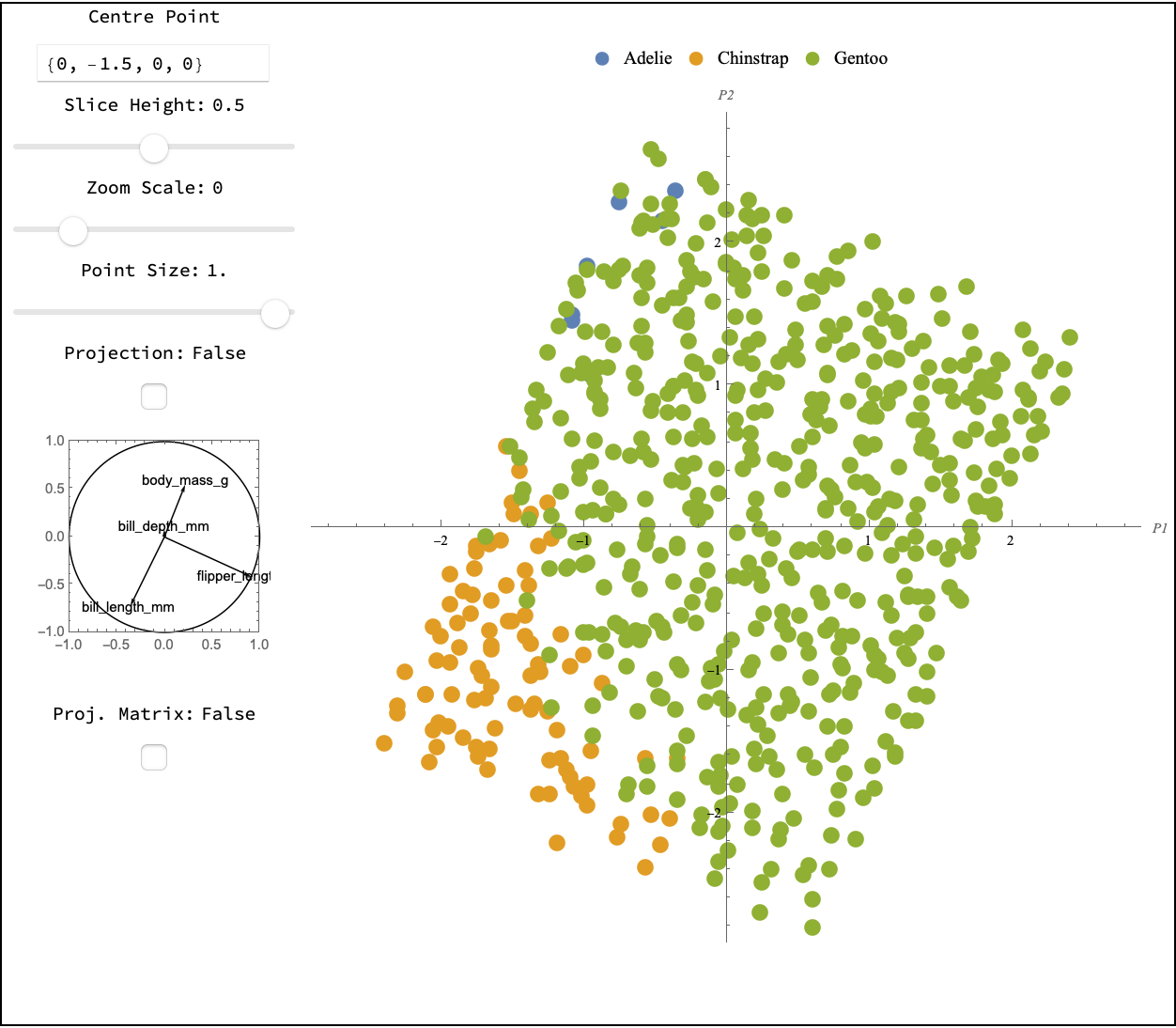}
\includegraphics[width=0.32\textwidth]{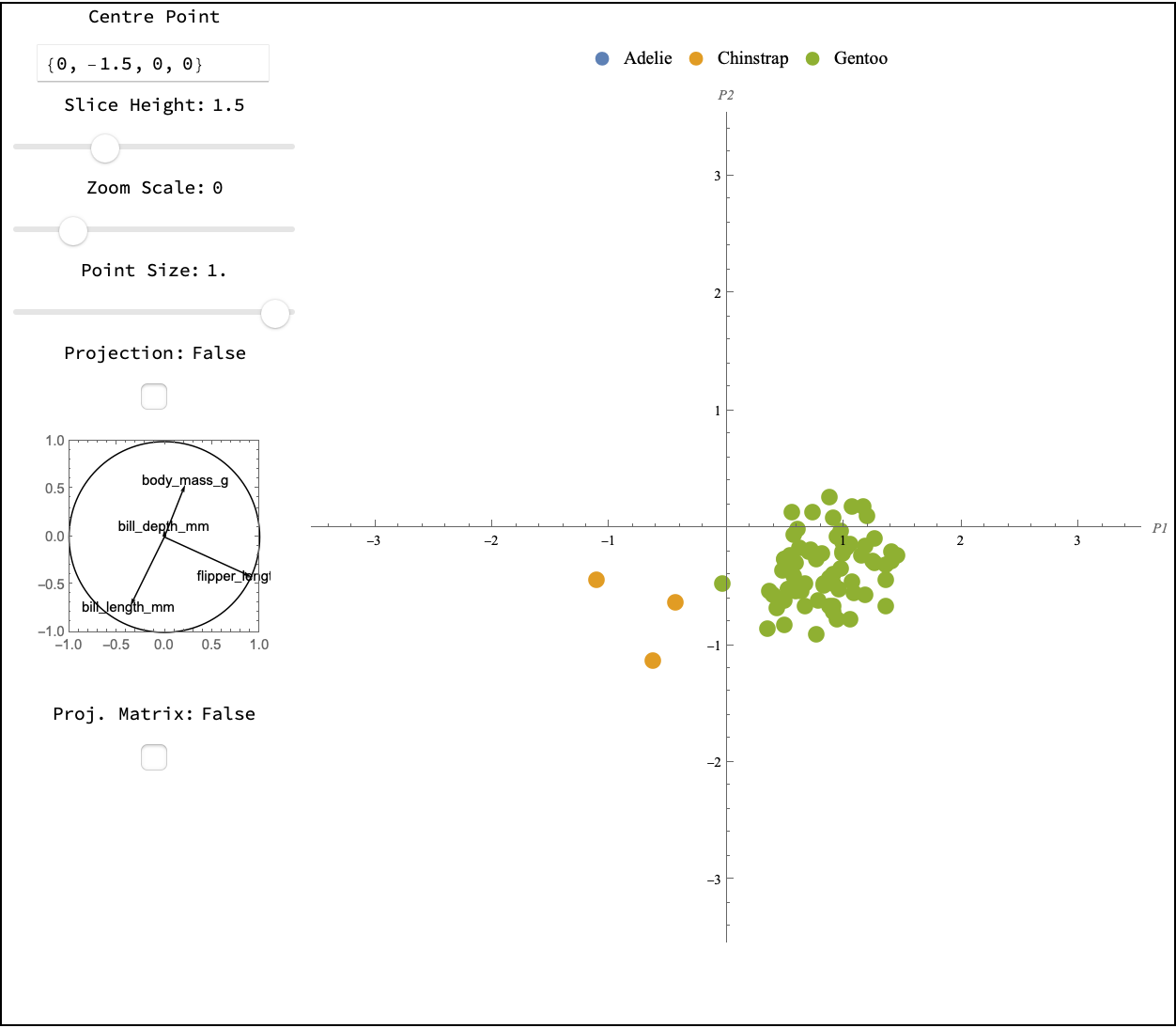}}
\caption{Shifting the slice center in the negative direction of bill depth produces regions that are mostly Gentoo (green). The two models have very different boundaries: the nonlinear potential of RF can be seen here where still some subspaces would predict to be Adelie and Chinstrap.}
\label{slice1m}
\end{figure*}

A more interesting comparison is found for \(\mathbf{c}^{-}\), thus the
slice localized towards low values of bill depth, shown in Figure
\ref{slice1m}. The RF model (left) predicts all three species within
this slice, with an interesting boundary for the third class (green,
Gentoo). On the other hand the LDA model (middle) predominantly predicts
the third class within the slice, this appears to be enforced through
the linear structure of the model. Looking finally at the thick slice
through the data we see that there are primarily observations from this
class within the slice we can conclude that the two models have filled
in the ``empty'' space (where we do not have any training observations)
in very different ways and according to what we might expect given the
model structure.

Finally it is interesting to compare the slice views to the projection
of the models seen in Fig. \ref{proj1} to better understand how the
boundaries change along the \texttt{bd} direction and where the
differences in the projections come from.

This example can be reproduced with the code in
penguins\_shifting\_slice\_center.nb and a run through is shown in the
second part of the video at \url{https://vimeo.com/747590472}.

\hypertarget{sec:discussion}{%
\section{Discussion}\label{sec:discussion}}

This short note has described new technology for manually interacting
with tours. The manual control of projection coefficients is most useful
for assessing variable importance to perceived structure, but can be
generally used for steering a viewer through high-dimensional space.
Changing the center of a slice manually enables exploring the space
orthogonal to a projection, specifically in the direction of a single
variable. This is a new tool, and as shown in the example, when used
with the slice tour can be very useful for understanding boundaries of
classifiers. This is also likely a useful method for examining
low-dimensional non-linear manifolds, and also functions of multiple
parameters.

Mathematica provided a useful sandbox to experiment with the ideas
presented in this paper. Most of the data, though, is first constructed
using R, especially the classification boundaries. If new technology for
interactive graphics becomes available in R it would be useful to create
a tighter coupling of models and visualization to allow exploring and
comparing fits.

\hypertarget{acknowledgements}{%
\section*{Acknowledgements}\label{acknowledgements}}
\addcontentsline{toc}{section}{Acknowledgements}

The authors gratefully acknowledge the support of the Australian
Research Council and the ResearchFirst undergraduate research program at
Monash University. The paper was written in \texttt{rmarkdown}
\citep{rmarkdown} using \texttt{knitr} \citep{knitr}. The scatterplot
matrix of the penguins data was produced by the GGally \citep{GGally}
package built on ggplot2 \citep{ggplot2} graphics. We thank the
Institute of Statistics, BOKU, for their hospitality while part of this
work was conducted.

\hypertarget{supplementary-material}{%
\section*{Supplementary material}\label{supplementary-material}}
\addcontentsline{toc}{section}{Supplementary material}

The source material and animated gifs for this paper are available at
\url{https://github.com/uschiLaa/mmtour}.

The supplementary materials include:

\begin{itemize}
\tightlist
\item
  The Mathematica source code defining the new functions in mmtour.wl.
\item
  Three Mathematica notebooks with the code used in the application
  (corresponding to the three subsections).
\item
  Appendix with additional details about manual controls, and
  Mathematica functions.
\item
  Animations illustrating the manual tour and slicing, matching the
  static figures in the paper. These are also available at
  \url{https://vimeo.com/747585410} and
  \url{https://vimeo.com/747590472}.
\end{itemize}

\appendix

\hypertarget{refinements-to-enforce-exact-position-1}{%
\section{Refinements to enforce exact
position}\label{refinements-to-enforce-exact-position-1}}

The problem with the new simple method (Algorithm 1) is that the precise
values for \(V_m\) cannot be specified because the orthonormalisation
will change them.

\hypertarget{adjustment-method-1}{%
\subsection{Adjustment method 1}\label{adjustment-method-1}}

\begin{figure}
\includegraphics[width=1\linewidth]{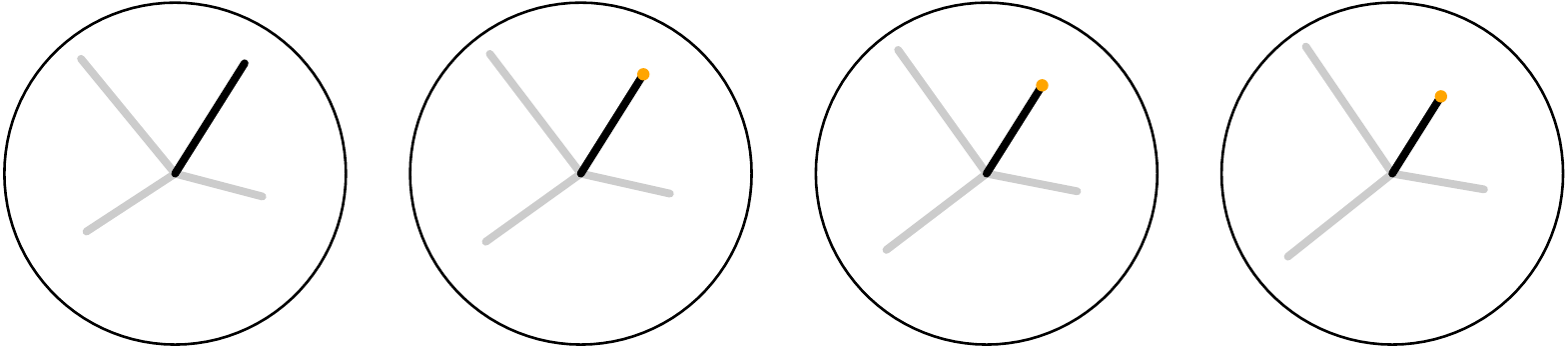} \caption{Manual controls with algorithm 2. The precise location of the axis is maintained.}\label{fig:othermethod}
\end{figure}

A small modification to algorithm 1 will maintain the components of
\(V_m\) precisely (Figure \ref{fig:othermethod}). It is as follows:

\begin{enumerate}
\def\labelenumi{\arabic{enumi}.}
\tightlist
\item
  Provide \(A\), and \(m\).
\item
  Change values in row \(m\), giving \(A^*\).
\item
  Store row \(m\) separately, and zero the values of row \(m\) in
  \(A^*\), giving \(A^{*0}\).
\item
  Orthonormalise \(A^{*0}\), using Gram-Schmidt.
\item
  Replace row \(m\) with the original values, giving \(A^{**}\).
\item
  For \(d=2\), adjust the values of \({\boldmath a}^{**}_{.2}\) using
\end{enumerate}

\[a^{**}_{j2}+\frac{a_{m1}a_{m2}}{p-1}, j=1, ..., p, j\neq m.\]

which ensures that

\[\sum_{j=1, j\neq m}^p a^{**}_{j1}a^{**}_{j2} + a_{m1}a_{m2} = 0.\]

If \(d>2\) the process would be sequentially repeated in the same manner
that Gram-Schmidt is applied sequentially to orthormalise the columns of
a matrix. If \(d=1\) no orthonormalisation is needed, and the projection
vector would simply need to be normalized after each adjustment.

\hypertarget{adjustment-method-2}{%
\subsection{Adjustment method 2}\label{adjustment-method-2}}

For \(d=2\) projections, the projection matrix is the sub-matrix \(A\),
of \(O\) (an orthonormal basis for the p-dimensional space), formed by
its first two columns as illustrated in Eq. (A1) . Whereas
orthonormality of the basis for the p-dimensional space is given by
\(e_i\cdot e_j=\delta_{ij},{i,j,=1,\cdots, p}\), orthonormality of the
projection matrix is expressed as
\(\sum_{k=1}^p a_{ki} a_{kj}=\delta_{ij}, ~{i,j=1,2}, ~{k=1,\cdots,p}\).
Movement of the cursor takes the two components \({a_{m1},a_{m2}}\) into
a selected new value \({a^*_{m1},a^*_{m2}}\). Although the motion is
constrained by \(a^{*2}_{m1}+a^{*2}_{m2}\leq 1\), this is not sufficient
to guarantee orthonormality of the new projection matrix. One possible
algorithm to achieve this is

\begin{enumerate}
\item Cursor movement takes ${a_{m1},a_{m2}}\to {a^*_{m1},a^*_{m2}}$, called $V_m$ above.
\item The freedom to change the components $a_{m3}\cdots a_{mp}$ (the columns of $O$ not corresponding to the projection matrix $A$) is used to select a new orthonormal basis as follows:
\begin{enumerate}
\item For row $m$ one chooses $a^*_{m3}=\sqrt{1-a^{*2}_{m1}-a^{*2}_{m1}},~a^*_{m,k>3}=0$
\item For other rows, $a_{i\neq m,j\geq3}$ random selections in the range $(-1,1)$ are made.
\item The Gram-Schmidt algorithm is then used to obtain an orthonormal basis taking $e_m$ as the first vector (which is already normalized), and then proceeding as usual $e_1\to e_1-(e_1\cdot e_m) e_m$, $e_1\to e_1/(e_1\cdot e_1)$, etc, resulting in $A^{**}$.
\end{enumerate}
\item this results in the orthonormal basis $O^*$ and a new projection matrix with $A^{**}_{m1}=a^*_{m1},~A^{**}_{m2}=a^*_{m2}$.
\end{enumerate}

The random completion of \(O\) outside the projection matrix provides an
exploration of dimensions orthogonal to the projection plane. However,
it renders projections in a way that is not continuous and may be
distracting. This can be alleviated by replacing the random completion
with a rule restricting the size of the jumps, for example in step (b)
the values \(a_{i\neq m,j\geq3}\) can be left unchanged before
orthogonalisation.

\begin{figure*}[ht]
\centerline{\includegraphics[width=0.9\textwidth]{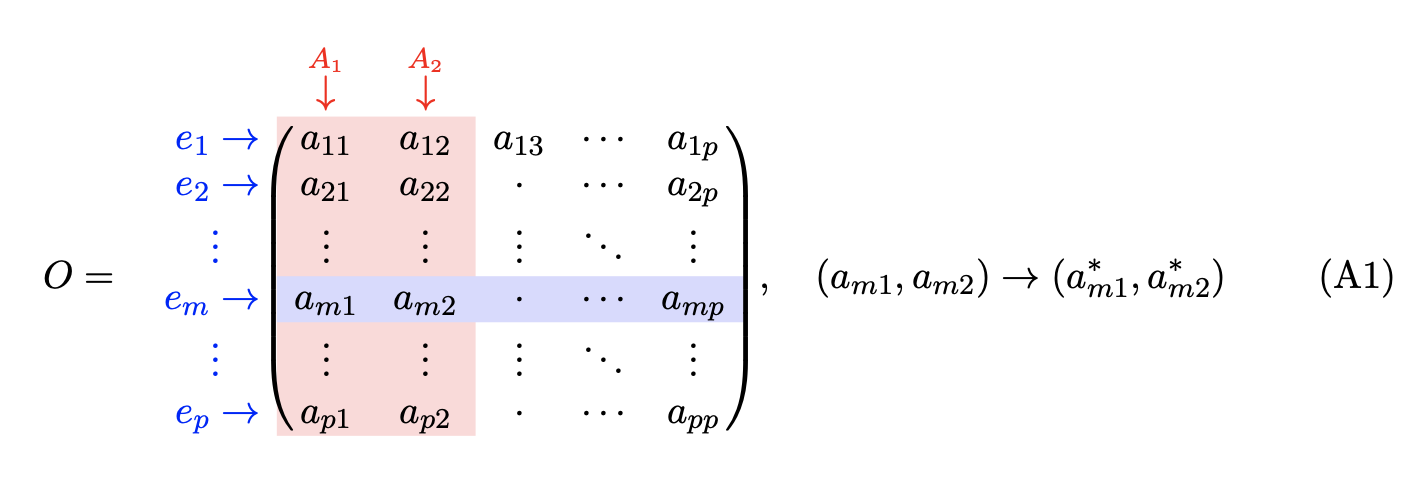}}
\end{figure*}

We could also take a similar approach without completing the basis,
through rotation within the \(d\)-dimensional hyperplane defined by the
new projection. In that case we would again start from \(A^*\) as above,
and then rotate the basis such that the first direction is aligned with
the selected variable \(m\). Applying Gram-Schmid within that basis
definition will not alter the direction for the first vector, thus the
direction of \(m\) remains fixed to what was selected with the cursor.
By adjusting the normalization procedure we can also ensure that the
lenght remains as selcted manually by the user.

\hypertarget{slice-center-guide}{%
\section{Slice center guide}\label{slice-center-guide}}

\begin{figure}
\includegraphics[width=1\linewidth]{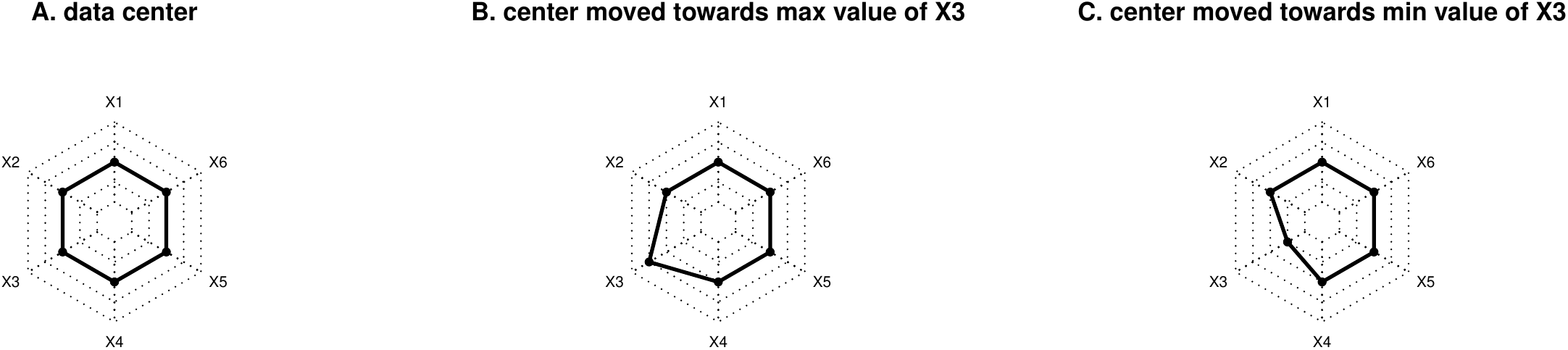} \caption{Visual guide for the slice center. Variable axes are displayed in polar coordinates, where the center corresponds to minimum value and outer end corresponds to maximum value. The position of the center corresponds to the dark polygon. If the center is at the center of the data, this will be displayed as a regular polygon (A), and plots B, C show it's position when moving the center along one axis.}\label{fig:anchornav}
\end{figure}

The slice center, \(c\), from Equation 1 in the main paper, is a point
in the \(p\)-dimensional space. By default, it is placed in the center
of the data (Figure \ref{fig:anchornav} A). Moving this along a variable
axis moves the slice outwards from the center of the data (Figure
\ref{fig:anchornav} B, C). The visual display of the center position is
a form of star plot as suggested useful in \citet{condviz2}, where the
position of the point is marked by a polygon on radial axes indicating
its position relative to the variable minimum to maximum values.

\hypertarget{software-details}{%
\section{Software details}\label{software-details}}

Here we describe the functions implemented in the Mathematica package
\texttt{mmtour.wl}. The main function and its arguments are given below:

\texttt{SliceDynamic{[}data,\ projmat,\ height,\ heightRange,\ legendQ(=1),\ flagQ(=1),\ colorFunc(=ColorData{[}97{]}){]}}

\begin{itemize}
\tightlist
\item
  \texttt{data}: A data matrix, potentially with grouping that has a
  label (string) and a flag (numerical) in the last two columns.
\item
  \texttt{projmat}: The initial projection matrix. It is possible to
  start with a random projection with the input ``random'' (including
  the quotes) in this entry field.
\item
  \texttt{height}: The initial height of the slice.
\item
  \texttt{heightRange}: The range of slice heights to be explored.
\item
  \texttt{legendQ}: A flag signaling whether the data matrix includes a
  column with group labels (1) or not (0).
\item
  \texttt{flagQ}: A flag signaling whether the data matrix includes a
  numerical flag for the groups.
\item
  \texttt{colorFunc}: specifies the color mapping for the different
  groups.
\end{itemize}

The first four arguments are required. As possible starting values we
suggest using a random input projection, and selecting slice height
parameters using the estimate given in Section 2.4.2 of the paper.

The function renders a manual tour with controls that allow the user to
navigate through different projections. For a given projection, a slider
permits the user to change the slice width within its range and a
separate box switches the view to a projection. The user can also change
the center point of slices; the size of the points rendered and the
scale of the plot through sliders. An additional box displays the
coordinates of the projection matrix explicitly.

This function can be used for ungrouped data by setting legendQ and
flagQ to 0. To remove a group from the visual display, the user can
choose specific colors for the entry colorFunc. For example,
\{Red,White,Green\} would display groups 1 (red) and 3 (green) while
making 2 invisible.

The package includes the additional functions not used in the example
application: \texttt{ProjectionPlot}, \texttt{Projected2DSliderPlot},
and \texttt{VisualiseSliceDyanmic}.

\texttt{ProjectionPlot} is very similar to \texttt{SliceDynamic}, except
it only displays projections. Using it instead of \texttt{SliceDynamic}
simplifies the display and is more efficient. The interface is slightly
different, the data file should not contain labels for groups, if these
are desired they are passed through the optional argument legendNames as
\{``group 1'', ``group 2'',\ldots\}, for example. The function and its
arguments are given below:

\texttt{ProjectionPlot{[}data,\ projmat,\ legendNames(=\{\}),\ colorFunc(=ColorData{[}97{]}){]}}

\begin{itemize}
\tightlist
\item
  \texttt{data}: A data matrix with one or more groups labeled by a flag
  (numerical, last column).
\item
  \texttt{projmat}: The initial projection matrix. It is possible to
  start with a random projection with the input ``random'' (including
  the quotes) in this entry field.
\item
  \texttt{legendNames}: an optional list of labels for the groups.
\item
  \texttt{colorFunc}: specifies the color mapping for the different
  groups.
\end{itemize}

\texttt{ProjectedLocatorPlot} displays the interactive controls slightly
differently. Several locator panes are displayed above the display of
the projected data and each one corresponds to a row in the projection
matrix. The updating behavior of the projection is the same as in
\texttt{ProjectionPlot}. The function and its arguments are given below:

\texttt{ProjectedLocatorPlot{[}data,\ projmat,\ legendNames(=\{\}),\ colorFunc(=ColorData{[}97{]}){]}}

\begin{itemize}
\tightlist
\item
  \texttt{data}: A data matrix with one or more groups labeled by a flag
  (numerical, last column).
\item
  \texttt{projmat}: The initial projection matrix. It is possible to
  start with a random projection with the input ``random'' (including
  the quotes) in this entry field.
\item
  \texttt{legendNames}: an optional list of labels for the groups.
\item
  \texttt{colorFunc}: specifies the color mapping for the different
  groups.
\end{itemize}

Finally \texttt{VisualiseSliceDynamic} allows to discern which points
exist inside the slice and outside it. This function differs from
\texttt{SliceDynamic} in that it looks at only one data set but displays
both the points inside and outside the slice. The display panel allows
the user to specify the size of both sets of points (inside and outside)
along with the slice height, center point and zoom level. The function
and its arguments are given below:

\texttt{VisualiseSliceDynamic{[}data,\ projmat,\ height,\ heightRange{]}}

\begin{itemize}
\tightlist
\item
  \texttt{data}: A data matrix containing only one group. In this case
  the data matrix should not contain column labels.
\item
  \texttt{projmat}: The initial projection matrix. It is possible to
  start with a random projection with the input ``random'' (including
  the quotes) in this entry field.
\item
  \texttt{height}: The initial height of the slice.
\item
  \texttt{heightRange}: The range of slice heights to be explored.
\end{itemize}

The points inside the slice are shown in black whereas those outside the
slice are shown in light blue, with labels indicating this.

\bibliographystyle{tfcad}
\bibliography{biblio.bib}

\end{document}